\documentclass[showpacs,amsmath,amssymb,superscriptaddress,preprintnumbers,nofootinbib,showkeys]{revtex4}

\begin{document}

\newcommand{\nablab}{{\mathop {\rule{0pt}{0pt}{\nabla}}\limits^{\bot}}\rule{0pt}{0pt}}

\title{Nonlocal extension of relativistic non-equilibrium thermostatics}

\author{Alexei S. Ilin}
\email{alexeyilinjukeu@gmail.com} \affiliation{Department of
General Relativity and Gravitation, Institute of Physics, Kazan
Federal University, Kremlevskaya str. 16a, Kazan 420008, Russia}

\author{Alexander B. Balakin}
\email{Alexander.Balakin@kpfu.ru} \affiliation{Department of
General Relativity and Gravitation, Institute of Physics, Kazan
Federal University, Kremlevskaya str. 16a, Kazan 420008, Russia}

\date{\today}

\begin{abstract}
Based on the formalism of nonlocal extension of the Israel-Stewart causal thermodynamics \cite{1} on the one hand, and on the formalism of the extended thermostatics \cite{2} on the other hand, we propose the new model of nonlocal relativistic non-equilibrium thermostatics for description of the static spherically symmetric  stellar systems. This nonlocal formalism operates with the pair of orthogonal four-vectors: one of them is the standard unit timelike medium velocity four-vector, the second one is the unit spacelike director. We derived the extended equation describing the profile of the non-equilibrium pressure, which can be indicated as the static analog of the Burgers constitutive equation known in classical rheology.
\end{abstract}
\maketitle

\section{Introduction}

The description of the neutron star structure involves into consideration the formalism of equations of state. We believe that the nonlocal rheological-type extension of the relativistic non-equilibrium irreversible thermodynamics and thermostatics is the most interesting trend in this direction, and in two recent works \cite{1,2} we have prepared the corresponding mathematical basis. In the work \cite{33} we have made the trial step towards the description of rheological-type equation of state for the neutron stars at zero temperature. In this note we suggest the new nonlocal model adapted for static spherically symmetric rheological systems. In this context we consider the heat-flux four-vector $q^k$ and the traceless part of the non-equilibrium pressure tensor $\Pi^{(0)}_{mn}$ to be vanishing. We search for the scalar part of the pressure tensor $\Pi$ as a function of the radial variable $r$.

\section{The formalism}

\subsection{Geometrical aspects of the theory}

We use the canonic line element for the model of static spherically symmetric spacetime
\begin{equation}
ds^2 = B(r)dt^2 - A(r)dr^2 - r^2\left({d\theta}^2 + \sin^2{\theta} {d\varphi}^2 \right) \,, \quad B(\infty) = 1 = A(\infty) \,.
\label{metric}
\end{equation}
The symmetry of this spacetime is described by the following Killing vectors:
\begin{equation}
\xi^j_{(0)} = \delta_0^j   \,, \quad \xi^j_{(1)} = \sin{\varphi}\delta_{\theta}^j {+} \cot{\theta} \cos{\varphi} \delta^j_{\varphi}\,, \quad \xi^j_{(2)} = \cos{\varphi}\delta_{\theta}^j {-} \cot{\theta} \sin{\varphi} \delta^j_{\varphi}\,, \quad  \xi^j_{(\varphi)} = \delta_{\varphi}^j \,.
\label{r30}
\end{equation}
One obtains from (\ref{r30}) and (\ref{metric}) that the squares of the Killing vectors give the following relationships:
\begin{equation}
\xi^2_{(0)} \equiv g_{kj} \xi^k_{(0)} \xi^j_{(0)} = B  \,,
\quad \xi^2_{(\varphi)} \equiv g_{kj}\xi^k_{(\varphi)}\xi^j_{(\varphi)} = - r^2 \sin^2{\Theta} \,,
\quad \xi^2_{(1)} + \xi^2_{(2)} + \xi^2_{(\varphi)} = - 2 r^2
 \,.
\label{squares}
\end{equation}

\subsection{The velocity four-vector and the director}

The velocity four-vector and its covariant derivative are, respectively, of the form
\begin{equation}
U^i = \delta^i_0 \frac{1}{\sqrt{B}} \,, \quad \nabla_k U_i = - \delta^0_k \delta^r_i \frac{B^{\prime}}{2\sqrt{B}} \  \Rightarrow DU_i \equiv U^k \nabla_k U_i = -  \delta^r_i \frac{B^{\prime}}{2B}  \,.
\label{18}
\end{equation}
The prime denotes the derivative with respect to the radial variable. For these quantities we deal with the vanishing expansion scalar  ($\Theta=0$), shear tensor ($\sigma_{ik}=0$), and vorticity tensor ($\omega_{ik}=0 $); the only acceleration four-vector $DU_i$ is non-vanishing. With (\ref{18}) the Eckart's version of thermodynamics predicts that, when the heat-flux four-vector $q^k$ vanishes \cite{Eckart}, the temperature $T(r)$ is predetermined to be of the following form:
\begin{equation}
q^i=0 \ \Rightarrow  \  \frac{1}{T} \nablab_k T = DU_k \ \Rightarrow  \ T(r)= \frac{T(\infty)}{\sqrt{B(r)}} \,.
\label{21}
\end{equation}
The director ${\cal R}^i$ can be defined as the unit spacelike  four-vector orthogonal to the velocity four-vector, which inherits the symmetry of the spacetime, i.e.
\begin{equation}
{\cal R}^i U_i =0 \,, \quad {\cal R}_i {\cal R}^i = -1 \,, \quad {\cal L}_{\xi_{(a)}}{\cal R}^i = \xi^k_{(a)} \partial_k {\cal R}^i - {\cal R}^k \partial_k \xi^i_{(a)} = 0 \,,
\label{r1}
\end{equation}
where ${\cal L}_{\xi_{(a)}}$ is the Lie derivatives along all the Killing vectors (\ref{r30}).
All these requirements are satisfied for the four-vector ${\cal R}^i = \delta^i_r \frac{1}{\sqrt{A}}$;  clearly, it possesses the following properties:
\begin{equation}
\nabla_k {\cal R}^i = \frac{1}{2\sqrt{A}}\left[\frac{B^{\prime}}{B} \delta^i_0 \delta_k^0 {+} \frac{2}{r} \left( \delta^i_{\theta} \delta_k^{\theta} {+} \delta^i_{\varphi} \delta_k^{\varphi} \right) \right]  \,, \quad \nabla_k {\cal R}^k = \frac{1}{2\sqrt{A}}\left[\frac{B^{\prime}}{B} {+} \frac{4}{r} \right] \,, \quad  {\cal R}^k \nabla_k {\cal R}^i = 0 \,,  \quad {\cal R}^k \nabla_k U^i =0 \,.
\label{r6}
\end{equation}
Using the analogy with the convective derivative $D \equiv U^k \nabla_k$ we introduce the directional derivative ${\cal D} \equiv {\cal R}^k \nabla_k$.
Finally, keeping in mind that there exists the divergence-free timelike four-vector $N^k = n U^k$, $\nabla_k N^k=0$, with the particle number density $n$, we introduce the new divergence-free spacelike four-vector $M^k = h {\cal R}^k$, $\nabla_k M^k =0$. The multiplier $h$ is considered to be constructed using the moduli of the Killing vectors (\ref{r30}) as follows:
\begin{equation}
h = \frac{h_0}{\xi_{(0)} \left[\xi^2_{(1)} + \xi^2_{(2)} + \xi^2_{(\varphi)}\right]} \ \Rightarrow  \nabla_k \left(h{\cal R}^k \right) = \frac{1}{r^2 \sqrt{AB}} \frac{d}{dr}\left[-\frac{h_0}{2} \right] =0 \,.
\label{r62}
\end{equation}

\subsection{Extension of the entropy flux four-vector}

The decomposition of the entropy flux four-vector $S^k$ can be organized, e.g., as follows:
\begin{equation}
S^k = S^k_{(\rm IS)} + h{\cal R}^k \left\{ {\cal D}^{-1}(\Pi f(T)) +  \frac12 \left[{\cal D}^{-1}\Pi \right]^2  + \frac12 \tau \Pi^2 \right\} \,,
\label{26}
\end{equation}
where the contribution $S^k_{(\rm IS)}$ appeared in the Israel-Stewart theory \cite{IS}. Here $\tau$ is a constant and $f(T)$ is some function of the temperature. The inverse operator ${\cal D}^{-1}$ is defined as
\begin{equation}
{\cal D}{\cal D}^{-1}\Pi= \Pi \ \Rightarrow \ {\cal D}^{-1}\Pi =
\int_{\infty}^r dr \Pi(r) \sqrt{A(r)}\,.
\label{266}
\end{equation}
For the static spherically symmetric model the entropy production scalar takes the form
\begin{equation}
\sigma = \nabla_k S^k = h \Pi \left[f(T) + {\cal D}^{-1}\Pi   + \tau {\cal D} \Pi \right] = \frac{\Pi^2}{9\zeta} \geq 0 \,,
\label{262}
\end{equation}
where $\zeta$ is some function of the temperature.
Thus, the non-equilibrium pressure $\Pi$ has to satisfy the integro-differential equation
\begin{equation}
\tau {\cal D} \Pi - \frac{\Pi}{9\zeta h} + f(T) + {\cal D}^{-1}\Pi = 0   \,.
\label{2680}
\end{equation}
The pure differential version of this equation is
\begin{equation}
\tau {\cal D}^2 \Pi - \frac{1}{9\zeta h} {\cal D} \Pi + \Pi \left[1- {\cal D}\left(\frac{1}{9\zeta h} \right)\right] + \frac{df}{dT} {\cal D}T = 0
  \,.
\label{268}
\end{equation}
We deal with the linear differential equation of the second order known as the Burgers constitutive equation \cite{Burgers}. The inhomogeneity of the temperature with the law ${\cal D}T = -\frac{T(\infty) B^{\prime}(r)}{2\sqrt{AB^3}}$ predetermines the features of the pressure distribution $\Pi(r)$; when $f(T)=const$ the equation (\ref{268}) admits the vanishing pressure $\Pi = 0$.

\section{Outlook}

The equation (\ref{268}) gives us the new equation describing the profiles of the non-equilibrium pressure $\Pi(r)$ in the model of static spherically symmetric object. The next step we have to do is to formulate the extended equation of hydrostatic equilibrium, which is the key element of the analysis of the star structure. We hope to analyze this problem in the nearest future.

\end{document}